%% file: main.tex
\documentclass[sigplan,twocolumn]{acmart}
\makeatletter
\def\@ACM@checkaffil{
    \if@ACM@instpresent\else
    \ClassWarningNoLine{\@classname}{No institution present for an affiliation}%
    \fi
    \if@ACM@citypresent\else
    \ClassWarningNoLine{\@classname}{No city present for an affiliation}%
    \fi
    \if@ACM@countrypresent\else
        \ClassWarningNoLine{\@classname}{No country present for an affiliation}%
    \fi
}
\makeatother

\renewcommand\footnotetextcopyrightpermission[1]{} 
\usepackage{tikz}
\usepackage{amsmath}

\usepackage{booktabs}

\usepackage{xcolor}
\usepackage{xspace}
\usepackage{pifont}

\usepackage{graphicx}

\usepackage{balance}

\usepackage{enumitem}

\usepackage{multirow}
\usepackage{makecell}

\usepackage{subcaption}
\usepackage{graphicx}
\usepackage{epstopdf}

\usepackage{tabularx}

\usepackage{ulem}

\def\Snospace~{\S{}}

\newcommand{\autorefsuffix}[2]{\hyperref[#1]{\autoref*{#1}#2}}

\definecolor{darkgreen}{rgb}{0.078,0.667,0.016}

\newcommand{\para}[1]{\noindent \textbf{#1 }}

\newcommand{\sys}{DynaServe\xspace}

\newcommand{\techgs}{fine-grained partition\xspace}
\newcommand{\techls}{SLO-aware batching\xspace}
\newcommand{\servera}{$\alpha$ server\xspace}
\newcommand{\serverb}{$\beta$ server\xspace}

\newcommand{\execmode}{Adaptive Partition and Scheduling\xspace}

\usepackage{listings}
\usepackage{soul}
\usepackage{hyperref}
\usepackage{amsmath}
\usepackage[linesnumbered,ruled,vlined]{algorithm2e}
\usepackage{pifont}

\begin{document}


\pagestyle{plain}

\title{\sys: Unified and Elastic Execution for Dynamic Disaggregated LLM Serving}

\author{Chaoyi Ruan$^\ast$}
\affiliation{%
  \institution{NUS}
}
\email{ruancy@comp.nus.edu.sg}

\author{Yinhe Chen$^\ast$}
\affiliation{%
  \institution{USTC}
}
\email{chenyh18@mail.ustc.edu.cn}

\author{Dongqi Tian}
\affiliation{%
  \institution{USTC}
}
\email{dongqitian@mail.ustc.edu.cn}

\author{Yandong Shi}
\affiliation{%
  \institution{USTC}
}
\email{yandongshi@mail.ustc.edu.cn}

\author{Yongji Wu}
\affiliation{%
  \institution{UCB}
}
\email{yongji.wu@berkeley.edu}

\author{Jialin Li}
\affiliation{%
  \institution{NUS}
}
\email{lijl@comp.nus.edu.sg}

\author{Cheng Li}
\affiliation{%
  \institution{USTC}
}
\email{chengli7@ustc.edu.cn}

\thanks{
$^\ast$Chaoyi Ruan and Yinhe Chen equally contributed to this work.
}

\begin{abstract}

\noindent LLM inference must meet strict latency SLOs (e.g., 100 ms P99 time-between-tokens) while maximizing goodput. Yet, real-world variability in prompt and response lengths skews compute‐intensive prefill and memory‐bound decode phases, making both colocated (even with chunked prefill) and disaggregated deployments unable to simultaneously deliver low tail latency and high throughput.

We introduce \sys, a high‐performance LLM serving system built atop vLLM that unifies and extends both paradigms for maximizing goodput under SLO constraints, when handling unbalanced and dynamic workloads. It relies on a \textit{micro-request} abstraction, which arbitrarily splits each request at any token boundary into at most two cooperating segments. A \textit{two-level scheduling} framework then balances micro-request load across unified GPU instances. The global scheduler rapidly selects per-request split points by considering both the request’s prefill/decode time ratio and the current load across GPU instances. The local schedulers on each GPU instance independently form SLO-aware batches, adjusting their composition in response to workload fluctuations, potential latency spikes and per-GPU under/over utilization. 
Finally, \sys uses chunked KV cache transfers to support cross-instance micro-request execution. 
On real-world traces, \sys boosts the overall serving capacity from 1.15$\times$ to 3.07$\times$, improves goodput by up to 1.91$\times$ and 1.61$\times$, and improves the performance by up to 60\% in a hybrid workload under SLO compared to state-of-the-art colocated and disaggregated baselines.

\end{abstract}

\maketitle

\input{sec/intro}

\input{sec/back}

\input{sec/motiv}

\input{sec/overview}
\input{sec/design}

\input{sec/eval}

\input{sec/relate}

\input{sec/conclusion}

\balance
\normalem
\bibliographystyle{plain}
\bibliography{ref}

\end{document}

%% file: sec/intro.tex
\section{Introduction}

\noindent Large language models (LLMs) have emerged as foundational components in modern applications ranging from code generation~\cite{chen2021codex,bavarian2022autoprompt}, customer‑support chatbots~\cite{thoppilan2022lamda,wulf2024explore}, and scientific‑computing assistants~\cite{culver2024scientific}.
An LLM serving system must handle high volume of concurrent inference requests, each imposing stringent latency SLOs to preserve user experience.
In particular, modern systems often target a 100 ms P99 time-between-tokens (TBT) bound to achieve fluid, real-time generation~\cite{zhong2024distserve}.
At the same time, service providers aim to maximize goodput (the number of requests served per second per GPU) since it directly translates to cost efficiency.
Striking a balance between tight latency SLOs and high goodput is thus both critical and challenging.

The LLM inference process naturally consists of two stages. The \textit{prefill} stage processes all input prompt tokens in parallel, generating the first output token and populating the KV cache. The \textit{decode} stage then produces subsequent tokens one-at-a-time, each step reloading the growing KV cache and accessing all prior token representations. Prefill workloads are compute-intensive, whereas decode is memory-bound. Because they share weights and KV cache, conventional serving systems typically colocate both phases on the same GPU instance to amortize resource usage~\cite{yu2022orca,agrawal2023sarathi,kamath2024pod}.

\begin{figure}[!t]
    \centering
    \includegraphics[width=0.40\textwidth]{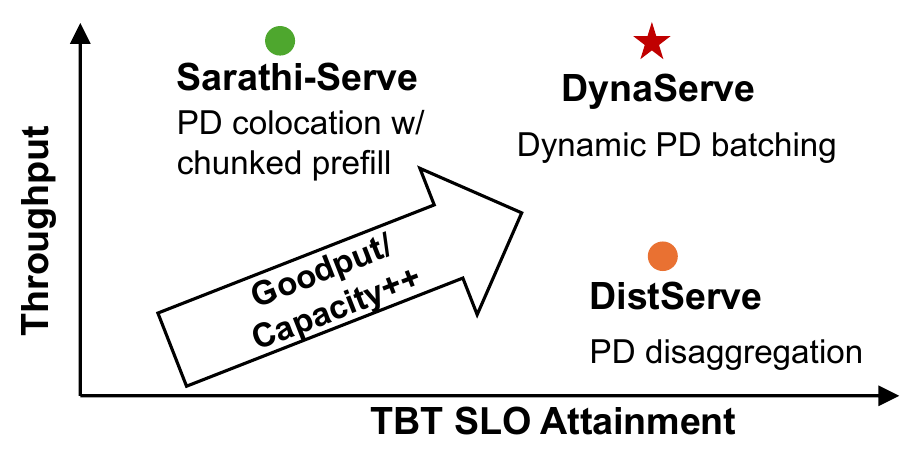}    
    \caption{
    Throughput vs. SLO attainment across serving architectures. PD colocation with chunked prefill reaches high throughput but violates the latency SLO. PD disaggregation satisfies the SLO but under-utilizes GPUs. \sys balances the two, advancing the frontier toward the top-right with higher goodput at guaranteed latency.}
    \label{fig:sec1.system.comparison}
\end{figure}

Colocation works well only when prefill and decode durations are balanced, which rarely holds in practice. As we show in Section~\ref{sect:motiv:workload}, real-world traces exhibit persistent and time-varying skew: some requests are heavily prefill-bound (long contexts), while others are decode-heavy (long continuations). This temporal imbalance causes interference: long prefill batches stall latency-sensitive decode steps, inflating TBT tail-latencies, while heavy decode phases can degrade prompt-processing responsiveness.

To mitigate interference, \textit{chunked prefill} extends the conventional colocation solution by splitting long prompts into smaller segments so decodes can interleave with prefill chunks~\cite{agrawal2023sarathi}. However, it only provides coarse latency bounds. \textit{PD disaggregation}~\cite{zhong2024distserve} assigns prefill and decode to separate GPU instances, eliminating interference entirely but often under-utilizing hardware due to mismatched stage loads. \autoref{fig:sec1.system.comparison} illustrates the latency-throughput trade-offs between the two paradigms, and their drawbacks are magnified under dynamic, unbalanced workloads.

In this paper, we present \sys, a new serving framework that unifies and extends colocation and disaggregation via Adaptive Request Partitioning and Scheduling (APS). Unlike coarse or prefill-stage-only splits, APS introduces the micro request abstraction, which is a contiguous span of tokens that may cover any mix of prefill and/or decode work. Each LLM request can be arbitrarily split at any token boundary into up to two micro requests. This fine-grained partitioning lets \sys tailor execution to each request’s compute profile, balancing load across GPUs and minimizing cross-stage interference to meet tight SLOs.

At the top level, \sys's global scheduler predicts decode length and stage-wise costs for each inference request. It then quickly searches the vast space of possible split token positions, often tens to thousands per request, to identify near-optimal micro-request partitions within just a few milliseconds. By leveraging lightweight cost models and a handful of probes, it balances per-request latency constraints against overall GPU load, then routes micro-requests in round-robin fashion to the unified GPU pool for fine-grained execution.

Within \sys, all GPU instances are equal and unified. Any instance can process any micro-request. A local scheduler on each GPU instance carefully composes incoming micro-requests into batches. For each batch, it tunes three key factors, namely, batch size, prefill-to-decode token ratio, and decode context length, to sustain high utilization under the 100 ms P99 TBT SLO.  This per-batch adaptation also takes place in an ultra-fast fashion and prevents stalls and idle cycles, even as workloads shift.

We implement \sys atop vLLM~\cite{vllm}, featuring a two-level scheduler with integrated runtime batching and partitioning support. Additionally, we introduce a chunk-based KV-transfer mechanism that efficiently manages the fine-grained KV cache transfers between unified GPU instances induced by micro-request partitioning. We evaluate \sys on A100 GPU clusters using real-world workloads such as BurstGPT~\cite{burstgptdata}, Azure Code~\cite{azuredata}, and others. \sys outperforms prior designs, with 1.15$\times$–3.07$\times$ and 1.09$\times$–1.67$\times$ higher serving capacity over PD colocation and PD disaggregation, respectively. It also achieves up to 1.91$\times$ and 1.61$\times$ higher goodput over PD colocation and PD disaggregation, respectively, and improves performance by 60\% in hybrid workload. Also \sys maintains high SLO attainment while achieving high performance. These results highlight \sys's adaptability and efficiency under diverse and dynamic traffic patterns.

In summary, this paper's contributions are:

\begin{itemize} [leftmargin=*, topsep=0pt]
    \item A comprehensive analysis and comparison of PD disaggregation and PD colocation, highlighting their respective advantages and disadvantages.
    \item The \execmode abstraction: we introduce a novel execution model in which each LLM request is split into cooperating \(\alpha\)/\(\beta\) micro-requests that can be executed across multiple GPUs in a dynamic manner.
    \item A dynamic scheduling framework: we design a two-level flexible scheduling that optimizes micro-request placement, batch composition, and partition ratios to meet latency SLOs while maximizing GPU utilization.
    \item Conduct a comprehensive evaluation of \sys with various real workloads. 
\end{itemize}

%% file: sec/back.tex
\section{Background and Motivation}
\label{sect:background}

\subsection{LLM Inference}
\label{sect:bg:inference}

The large language models (LLMs) inference generates text autoregressively, predicting one token at a time based on previously generated ones~\cite{vaswani2017attention}. Typically, LLM inference consists of two distinct stages, where a \textit{prefill} stage is followed by a \textit{decode} phase. 
In the prefill stage, the model processes the entire input prompt in a single forward pass, which computes the key-value (KV) cache for all tokens and generates the first output token. Then, the decode stage generates the remaining output tokens one by one, each using the accumulated KV cache from previous steps, until a stopping condition is met (e.g., token limit or stop word).

As a standard optimization technique, KV cache~\cite{kwon2023efficient} accelerates the LLM inference process by storing computed KV tensors in GPU memory, eliminating per-decode step recomputations. The initial input tokens’ KV cache is computed and loaded into the cache in bulk by the prefill stage. In contrast, at each step of the decode stage, the newly generated token’s KV tensors are incrementally appended to the cache, enabling efficient reuse in subsequent steps.

It is well-known that the computational characteristics of the prefill and decode stages differ drastically: the former is compute-intensive, while the latter is memory-bound and exhibits low arithmetic intensity. Since all input prompt tokens are available up front, they can be processed in parallel during the prefill stage, achieving high GPU utilization. In contrast, the decode stage emits one token at a time, serializing work on the GPU and leaving many compute units idle. The relative intensity of these two phases strongly depends on the lengths of the input and output sequences.

\subsection{LLM Serving Systems and Optimizations}
\label{background:systemsupport}

\begin{figure}[!tb]
    \centering
    \includegraphics[width=0.5\textwidth]{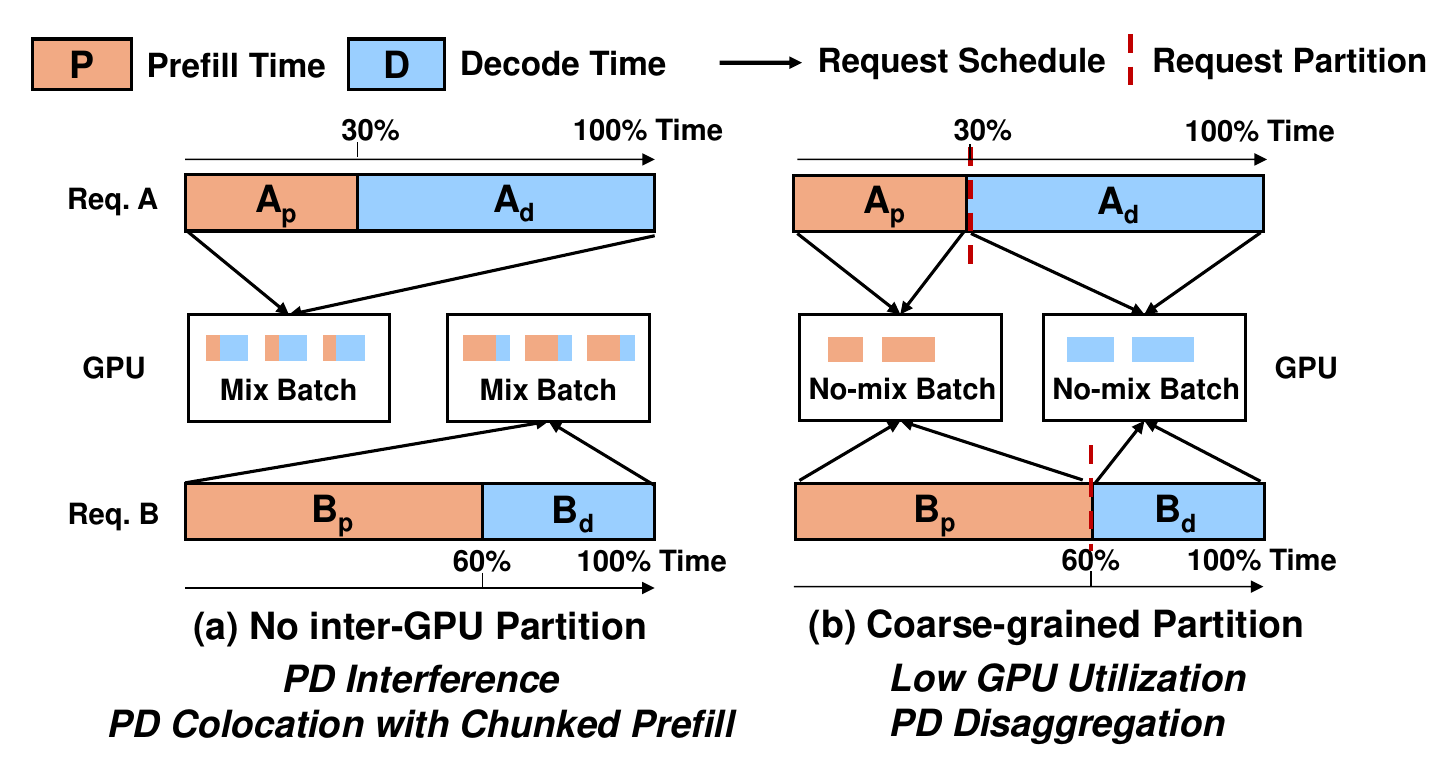}
    \caption{Partition and scheduling strategies in LLM serving. 
(a) PD colocation, applying chunked prefill to reduce interference. (b) coarse-grained PD disaggregation, which avoids interference but leads to GPU underutilization.}
    \label{bg:archi}
\end{figure}

Real-time online LLM services must simultaneously handle concurrent requests while meeting strict SLOs.
Key latency metrics include TTFT (time to first token) for prefill speed and TBT (time between tokens) for decode efficiency.
The sum of TTFT and the cumulative TBT determines overall request latency;
a bottleneck in either phase directly degrades user experience.
Overall throughput or GPU utilization, driven by concurrency, is another major metric for mainstream serving systems.
Since latency and throughput often conflict, most system optimizations revolve around balancing these objectives.
Our goal is to maximize LLM serving goodput~\cite{zhong2024distserve}, defined as the number of useful output tokens generated per second under a latency SLO. We begin by reviewing two common serving architectures (\autoref{bg:archi}) and their approaches to the latency–throughput trade-off.

\noindent\textbf{PD colocation (no inter-GPU partition).}
The simplest architecture runs both prefill and decode phases of a serving request on the same instance (\autoref{bg:archi}-(a)).
Colocation simplifies scheduling and enables straightforward continuous batching~\cite{yu2022orca}.
But since each instance handles prefill and decode phases of different requests concurrently, the scheme creates prefill-decode interference:
Heavy prefill workloads (especially for long contexts) may stall latency-sensitive decode steps, leading to long tail latencies between tokens.
To mitigate PD interference, recent systems like Sarathi-Serve~\cite{agrawal2024taming} and POD Attention~\cite{kamath2024pod} introduce \textit{chunked prefill}.
As illustrated in \autoref{bg:archi}-(a), the approach breaks an input prompt into smaller segments, each interleaved with decode execution.
POD Attention further fuses prefill and decode into a single GPU kernel, improving their overlap and GPU utilization.
However, effectiveness of chunked prefill is highly workload-dependent.
Only when the decode phase dominates (e.g., reasoning tasks where outputs can be 10$\times$ longer than inputs), can chunked prefill maintain acceptable tail latency as the shorter prefill stage causes minimal contention.
Furthermore, their latency and throughput are sensitive to the chunk size hyperparameter.
Mainstream solutions~\cite{vllm,sglang} choose a static, coarse-grained chunk size for each instance; unfortunately, the approach falls short when facing dynamic workloads.

\noindent\textbf{PD disaggregation (coarse-grained partition).}
To eliminate interference, PD disaggregation (\autoref{bg:archi}-(b)) physically separates prefill and decode stages onto dedicated GPU pools.
Systems such as DistServe~\cite{zhong2024distserve}, Mooncake~\cite{qin2024mooncake}, and Splitwise~\cite{patel2024splitwise} apply such an architecture and tune each phase independently.
Physical isolation enables PD disaggregation to offer more stable tail latencies.
The architecture achieves optimal throughput when prompt and output workloads are well-balanced, i.e., the prefill and decode stages have roughly a 1:1 compute ratio.
However, any imbalance in the workload, such as when prompt lengths increase or output lengths shrink, creates uneven resource utilization across the two GPU pools, resulting in resource fragmentation and degraded overall efficiency.
This sensitivity to workload skew presents a key challenge for disaggregated designs in real-world serving where prompt/output patterns are highly dynamic.

\subsection{Unbalanced and Dynamic Workloads}
\label{sect:motiv:workload}

\begin{figure}[!t]
    \centering
    \includegraphics[width=0.45\textwidth]{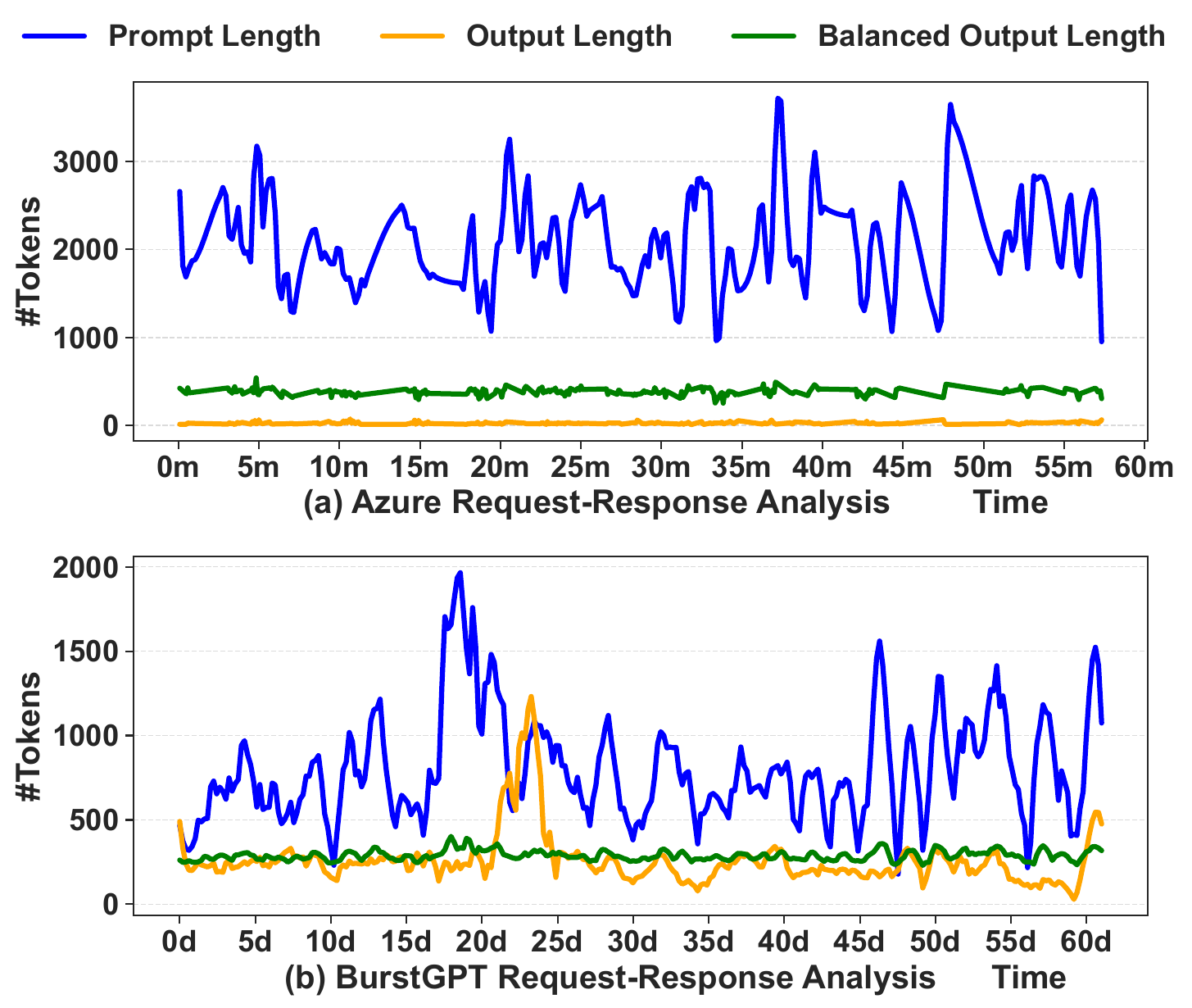}
    \caption{Prompt and output token lengths distribution. The blue line shows prompt length, and the orange line shows output length. The green line indicates the balanced output length, where decode time equals prefill time.}
    \label{fig:workload-distribution}
\end{figure}

We now examine two real-world request traces, Azure Code~\cite{patel2024splitwise} and BurstGPT~\cite{wang2024burstgpt}, to understand the time‐varying imbalance between the compute demands of the prefill and decode stages in LLM inference.~\autoref{fig:workload-distribution} shows the per-minute prompt token count (blue line) and response token count (yellow line) of each workload.
In each graph, we also plot a ``balanced'' decode curve (green), which shows the number of output tokens whose decode time would exactly match the prefill time. 
This calculation is done using measured prefill throughput on the same A100 GPU.
In regions of the graph where the yellow line exceeds the green curve, decode demands more GPU time than prefill; otherwise (yellow line falls under the green curve), prefill dominates.

In the Azure Code workload, the prompt curve consistently surpasses both the response and the balance lines.
This indicates that prefill tasks in this workload require more GPU time than decode and that decode tasks contribute only a small fraction of the total inference cost.
The trend also persists throughout the trace, revealing a persistent prefill‐heavy profile of the workload.

In contrast, BurstGPT exhibits wide swings in both prompt and response volumes.
There are extended regions where the response line (yellow) exceeds the balanced curve, indicating decode‐heavy periods, followed by regions of prefill dominance.
The trace also shows rapid fluctuations between the two types of regions, underscoring the presence of high temporal variance in real‐world workloads.
Similar dynamics also appear at scale.
For instance, Alibaba’s Infinite-LLM reports context lengths ranging from a few tokens to more than two million in production~\cite{lin2024infinitellmefficientllmservice}.

%% file: sec/motiv.tex
\subsection{Issues with Existing Techniques}
\label{sec:motiv:benchmark}

\begin{table}[!tbp]
\centering
\caption{GPU compute, KV cache usage, and inter-token latency when serving Qwen-2.5-14B~\cite{qwen-14b} on two A100 GPUs (G1 and G2) under PD disaggregation and colocation. Request rates are tuned to saturate the GPUs. For disaggregation, both prefill and decode instances are shown; for colocation, only the averaged metrics are reported due to symmetry.}
\label{tab:performance_comparison}
\resizebox{0.48\textwidth}{!}{
\begin{tabular}{@{}l@{\hskip 2pt}l@{\hskip 6pt}cccccc@{}}
\toprule
\textbf{Metric} &  & 
\multicolumn{2}{c}{\textbf{P-8192, D-32}} & 
\multicolumn{2}{c}{\textbf{P-2048, D-512}} &
\multicolumn{2}{c}{\textbf{P-219, D-1467}} \\
\cmidrule(lr){3-4} \cmidrule(lr){5-6} \cmidrule(lr){7-8}
& & \textbf{Disagg.} & \textbf{Coloc.} & \textbf{Disagg.} & \textbf{Coloc.} & \textbf{Disagg.} & \textbf{Coloc.} \\
\midrule
\multirow{2}{*}{\textbf{MFU (\%)}} & \textbf{G1} & 43.24 & 38.94 & 30.59 & 21.26 & 2.09 & 13.98 \\
                           & \textbf{G2} & 0.19 & 38.94 & 7.93 & 21.26 & 14.34 & 13.98 \\
\midrule
\multirow{2}{*}{\textbf{HBM Usage (\%)}} & \textbf{G1} & 4.83 & 39.67 & 1.17 & 94.23 & 0 & 95.2 \\
                                 & \textbf{G2} & 5.77 & 40.03 & 94.83 & 95.7 & 96.27 & 95.33 \\
\midrule
\textbf{p50-TBT (ms)} &     & \textbf{22.70} & 309.68 & \textbf{47.00} & \textbf{46.86} & \textbf{50.63} & \textbf{47.99} \\
\textbf{p99-TBT (ms)} &     & \textbf{58.09} & 352.93 & \textbf{65.11} & 336.81 & \textbf{74.47} & 162.67 \\
\textbf{Throughput (rps)} &     & 0.83 & \textbf{1.49} & 2.56 & \textbf{2.83} & 1.68 & \textbf{2.86} \\
\textbf{Attainment (\%)} &     & \textbf{100.00} & 1.73 & \textbf{99.83} & 84.09 & \textbf{99.95} & 94.46 \\
\bottomrule
\end{tabular}
}
\end{table}

How well do prior serving architectures handle real-world LLM workloads?
To answer this question, we benchmark both PD colocation with chunked prefill and PD disaggregation, using the imbalanced, time-varying workloads described above. 
We serve Qwen-2.5-14B on two NVIDIA A100 GPUs.
In the colocation setup, both GPUs run identical model replicas with round-robin request routing.
We use a 2048-token chunk size (default in vLLM) to balance prefill and decode execution.
In the disaggregated setup, we apply a 1P+1D configuration, where one GPU handles all prefill operations and the other handles decode, fully isolating the two stages.

For workloads, we select three representative request shapes based on Azure Code, BurstGPT (as discussed in \autoref{sect:motiv:workload}), and Mini Reasoning~\cite{mini-reasoning}.
These shapes reflect commonly observed workload patterns: (1) a long prompt with a short output (plen=8192, dlen=32), (2) a balanced input-output case (plen=2048, dlen=512), and (3) a short prompt with a long output (plen=219, dlen=1467).
These configurations capture both general-purpose and reasoning-centric scenarios.
For all cases, request rates are adjusted to saturate the system capacity, ensuring a meaningful comparison of throughput, latency, and resource efficiency under pressure.

We begin our analysis by examining how well each system can sustain latency under SLO.
\autoref{tab:performance_comparison} summarizes median (P50) and tail (P99) TBT for the three request shapes, compared to a 100 ms TBT SLO as defined by DistServe~\cite{zhong2024distserve}.
Tail latency of colocation with chunked prefill violates the SLO across all workloads: P99‐TBT exceeds 330 ms for both the long‐prompt (P = 8192, D = 32) and balanced (P = 2048, D = 512) workloads;
even in the reasoning‐style short‐prompt case (P = 219, D = 1467) where prefill interference should be minimal, tail latency still reaches above 160 ms.
Notably, the median latency in the long‐prompt scenario also violates the SLO, with P50‐TBT rising above 300 ms.
In contrast, PD disaggregation holds P50‐TBT below 60 ms across all traces and maintains P99‐TBT under 100 ms, consistently satisfying the latency SLO.

Next, we shift our focus to throughput and GPU resource utilization.
\autoref{tab:performance_comparison} reports the model FLOPs utilization (MFU), the HBM usage percentage, and the overall request per second (rps).
PD disaggregation shows pronounced resource utilization imbalance between the two GPUs: The prefill device (G1) exhibits high MFU but minimal HBM occupancy, while the decode device (G2) observes the opposite pattern.
For example, under the long‐prompt workload (P = 8192, D = 32), G1 achieves 43.2\% MFU with only 4.8\% HBM usage, while G2 delivers a mere 0.2\% MFU alongside 5.8\% HBM utilization.
The short‐prompt, long‐decode case (P = 219, D = 1467) leaves G1 largely idle while G2’s memory almost saturates.
When serving the balanced workload (P = 2048, D = 512), G1 underuses memory and G2 underuses computation.
All of these observations indicate poor overlap between compute and memory demands.
The colocation architecture, on the other hand, achieves balanced resource utilization across devices.
Consequently, disaggregation yields only 0.83 requests per second (rps) versus 1.49 rps in colocation (a 44\% drop) for long prompts, and suffers 41\% and 10\% throughput drops in the short‐prompt and balanced cases, respectively.

Together, these findings expose a fundamental tension: Prior serving designs fall short when the request pattern deviates even modestly from their ideal workload assumptions --- decode-heavy workloads for colocation and balanced workloads for disaggregation.
Yet, as our benchmarks show, even in workloads where prefill tasks are short or prompt/output lengths are numerically balanced, the two approaches suffer from either latency spikes or throughput loss.
This motivates us to design a more adaptive serving framework that can dynamically partition and schedule work at fine granularity, aligning GPU resource allocation with each request’s evolving compute profile.

%% file: sec/overview.tex
\section{\sys Overview: Approach and Challenges}
\label{sec:overview}

\noindent Our benchmark results in \autoref{sec:motiv:benchmark} show a latency-throughput trade-off in existing serving architectures when handling real-world workloads.
In this section, we give an overview of \sys, a new serving framework designed to efficiently handle dynamic and imbalanced LLM workloads.

\subsection{Design Overview}
\label{sect:overview:archi}

\begin{figure}[!t]
    \centering
        \includegraphics[width=0.45\textwidth]{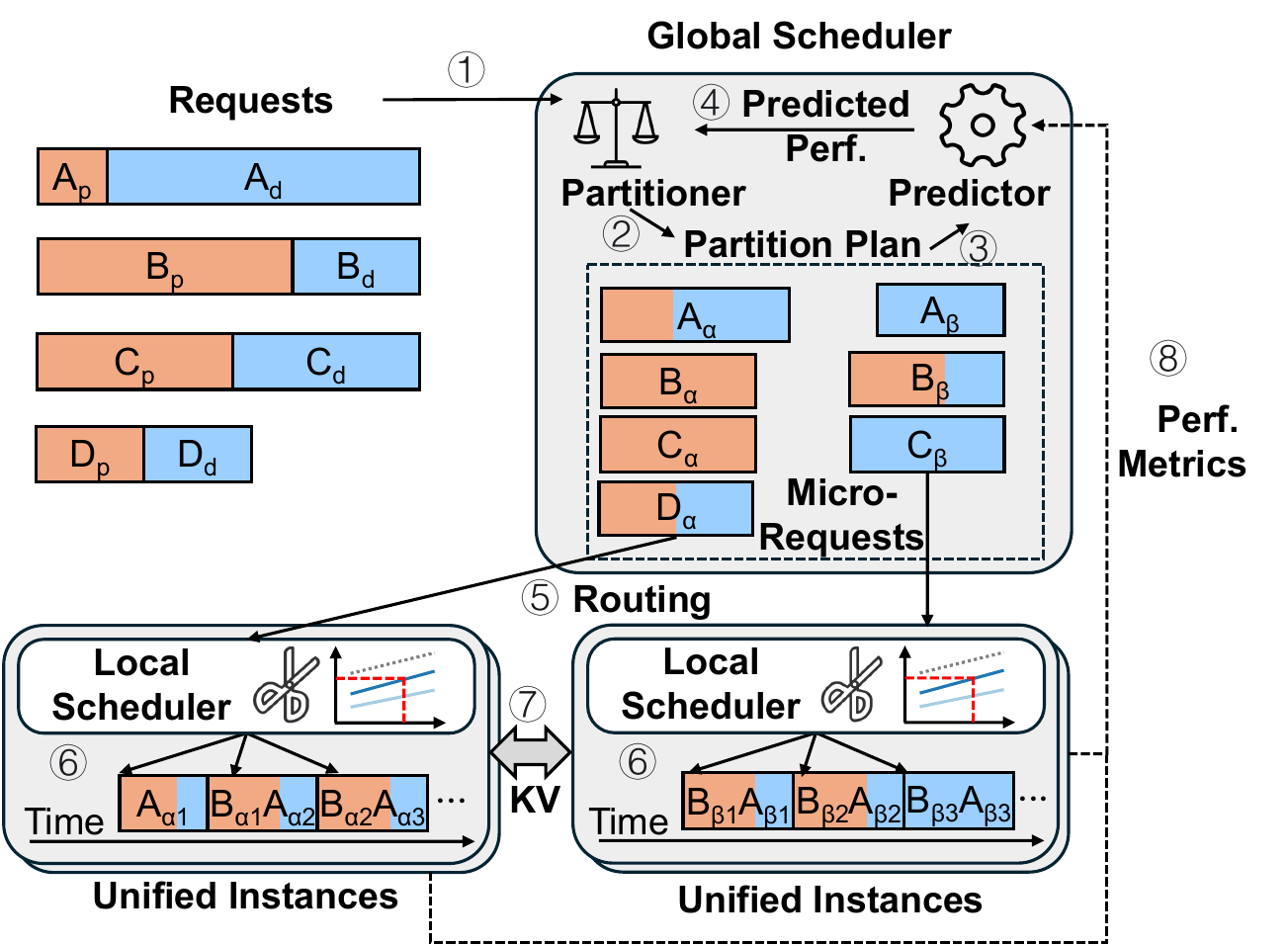}
    \caption{The overall architecture of \sys features unified GPU instances executing partitioned micro-requests, guided by a two-level APS mechanism for improved SLO attainment and resource utilization. Orange and blue colors denote the prefill and decode stages of each request, respectively.}
    \label{fig:overview}
\end{figure}

The key design principle in \sys is \textit{Adaptive Request Partition and Scheduling} (APS).
As shown in \autoref{fig:overview}, APS proposes a new micro-request abstraction that allows partitioning the prefill/decode phases of a serving request at \textit{any} token position.
Partitioning and dispatching decisions are centrally made by a global scheduler based on collected latency and resource utilization metrics. 
Finally, a runtime system local to each GPU server schedules, batches, and executes the assigned micro-requests to maximize serving throughput while guaranteeing latency SLO.

\noindent\textbf{Micro-request abstraction.}
Each serving request $r$ can be described by its prompt length $P$ and a decode length $D$;
we estimate the decode length using prior length-prediction methods~\cite{jin2023s,qiu2024efficientinteractivellmserving}.
This results in a total logical length of $L = P + D$.
We associate each request with a splitting point $s$ between 0 and $L$, dividing the request into two \textit{micro-requests}: $r^{\alpha}$ (tokens 1…$s$) and $r^{\beta}$ (tokens $s{+}1$…$L$).
When $s$ is at the request boundary (0 or $L$), one of the micro-requests ($r^{\alpha}$ or $r^{\beta}$) is empty, i.e., no partitioning.
A \textit{micro-request} is a contiguous span of tokens, representing either prefill, decode, or a mixture of both.
The abstraction makes APS more flexible than prior architectures:
PD colocation partitions a request only within the prefill stage, while PD disaggregation splits a request strictly between prefill and decode.
APS, on the other hand, can split a request at any token position.

The micro-request abstraction enables APS to adapt smoothly to workload variations.
For example, when the system is underutilized or the prompt is short, APS may avoid partitioning altogether, similar to a colocated engine by routing the full request to one GPU instance (see request $D_\alpha$ in \autoref{fig:overview}).
When prefill and decode are balanced under high load, APS may adopt a PD disaggregated configuration (see requests $C_\alpha$ and $C_\beta$).
Beyond these, APS supports hybrid splits, e.g., merging early prefill with partial decode (request $A_\alpha$), or offloading part of prefill to run alongside decode on another GPU (request $B_\beta$).

Conceptually, APS operates within a \textit{generalized} request partitioning space;
prior PD colocation and disaggregation are mere special cases within this broader execution space.
The additional partitioning options offer APS more flexibility in navigating the utilization-latency space while being adaptive to workload dynamism.

\para{Global scheduler.} 
All serving requests are first handled by the centralized global scheduler (\ding{172}).
The scheduler has two responsibilities.
First, it determines the optimal splitting point $s$ of each request.
The scheduler does so by searching a candidate partition ratio $\phi \in [0,1]$ that offers the best predicted performance (\ding{173}-\ding{175}), and calculates the splitting point $s$ as $\lceil \phi L \rceil$.
\sys divides the sequence into two micro-requests: $r^{\alpha}$ (tokens 1…$s$) and $r^{\beta}$ (tokens $s{+}1$…$L$).

Second, the scheduler routes the partitioned micro-requests to executors based on current system load and latency SLOs (\ding{176}).
It runs a lightweight performance predictor, which collects periodic runtime statistics (\ding{179}) from each GPU, including compute/HBM memory utilization, and request queue.
These metrics reflect current workload pressure, allowing the predictor to estimate the performance impact of placing $r^\alpha$ and $r^\beta$ on different instances.
The scheduler selects a placement plan (micro-request to executor) that minimizes load imbalance and maximizes throughput, while meeting the request’s SLO.

\para{Unified execution instances.}
A unified execution instance runs micro-requests dispatched by the global scheduler on its GPUs.
Given a set of buffered requests, a local scheduler builds token-level batches via an SLO-aware control mechanism (\ding{177}).
The batching protocol enforces two constraints: It should respect the HBM limit of the resident GPUs and it keeps the p99 TBT latency below the SLO.
The local scheduler also adapts to changes in workload and observed request latency.
It widens the batches and bumps throughput when request workload is moderate; if latency approaches the SLO, it reconfigures per-batch composition to reduce the prefill-decode interference.
When the micro-requests of an LLM request span two execution instances, the instances exchange the required KV cache blocks over RDMA (\ding{178}).

\subsection{Design Challenges}
\label{sect:overview:challenges}

Although APS can adapt on the fly to meet SLOs and boost performance, unlocking its full potential requires overcoming three key system challenges.

\para{\textit{Challenge 1:} dynamic request slicing and scheduling complexity.} 
Each incoming request exposes dozens to thousands of valid split points, and each resulting micro-request can be routed to multiple unified instances. As a result, the scheduling search space for $n$ requests of lengths $l_1, \dots, l_n$ grows exponentially as $\prod l_i$. The global scheduler must navigate this enormous space within a few milliseconds and with limited foresight: a split that appears balanced now may lead to overload moments later as new traffic arrives. These challenges necessitate fast, predictive cost models to estimate benefits and efficiently identify near-optimal splits while respecting per-request SLOs.

\para{\textit{Challenge 2:} fine-grained batch control under SLOs.} 
Unified instances execute a continuous sequence of batches, each containing a dynamic mix of prefill and decode tokens. This composition varies over time as micro-requests are assigned by the global scheduler, leading to fluctuating workload characteristics that must be handled efficiently at the batch level. Effective batch composition depends on tuning two key factors: the \textit{prefill-to-decode token ratio}, which governs interference between the two stages, and the \textit{context lengths} of decode tokens, which directly impact per-token latency. Additionally, the total number of tokens per batch influences overall GPU utilization, making batching a multi-objective optimization task. To meet latency SLOs while maintaining high throughput, the local scheduler must make fast, accurate decisions when forming each batch. Unlike prior systems that rely on static chunk sizes to loosely bound latency, our setting demands fine-grained control over batch composition in response to changing request mixes, making the problem substantially more complex.

\para{\textit{Challenge 3:} frequent cross-instance KV cache transfers.} 
APS with micro-requests leads to more frequent and fine-grained KV cache transfers between unified GPU instances compared to coarse-grained PD disaggregation, as it permits splitting at arbitrary points within either the prefill or decode stages. For example, in a reasoning task from our experiments, a micro-request that merges a decode segment with the prefill results in 3$\times$ more token cache transfers than standard disaggregation, significantly increasing KV cache volume. To prevent these transfers from becoming a bottleneck, \sys must move KV cache efficiently to avoid stalling subsequent computation.

%% file: sec/design.tex
\section{\sys's Design}

\noindent This section presents \sys's two-level scheduling architecture. Our ultimate goal is to maximize overall resource utilization while satisfying the per-request token-by-token (TBT) service-level objective (SLO). To achieve this, we jointly optimize two key decisions: the per-request partition ratios ($\phi_r$) and the per-instance dynamic batch composition. 

To improve resource utilization, \sys balances workloads across all GPU instances. When execution time of each instance differs significantly, it indicates that some instances are idle and underutilized, leading to a reduction in overall throughput. Thus, in~\autoref{sec:globalsched}, \sys uses execution time to estimate each instance’s load and dynamically adjusts the partition ratio $\phi_r$ so that execution time remains balanced across instances.
Within each instance, \sys must also maintain both high efficiency and low latency. As we will detail in~\autoref{sec:localsched}, prefill length within a batch introduces a fundamental trade-off between GPU utilization and request latency. To manage this trade-off, \sys dynamically adjusts batch composition, extending prefill lengths when latency budgets allow to improve compute efficiency while still meeting the TBT SLO. This adjustment relies on a latency profiling table that estimates batch latency and is continuously updated at runtime with new measurements to refine its predictions.

Finally, \autoref{sec:chunk-transfer} describes runtime support mechanisms that enable unified micro-request execution and efficient key–value state transfer between prefill and decode stages.

\subsection{Global: Request Partition and Routing}
\label{sec:globalsched}
\label{sec:profiler}

\begin{figure}[!t]
    \centering
    \includegraphics[width=0.45\textwidth]{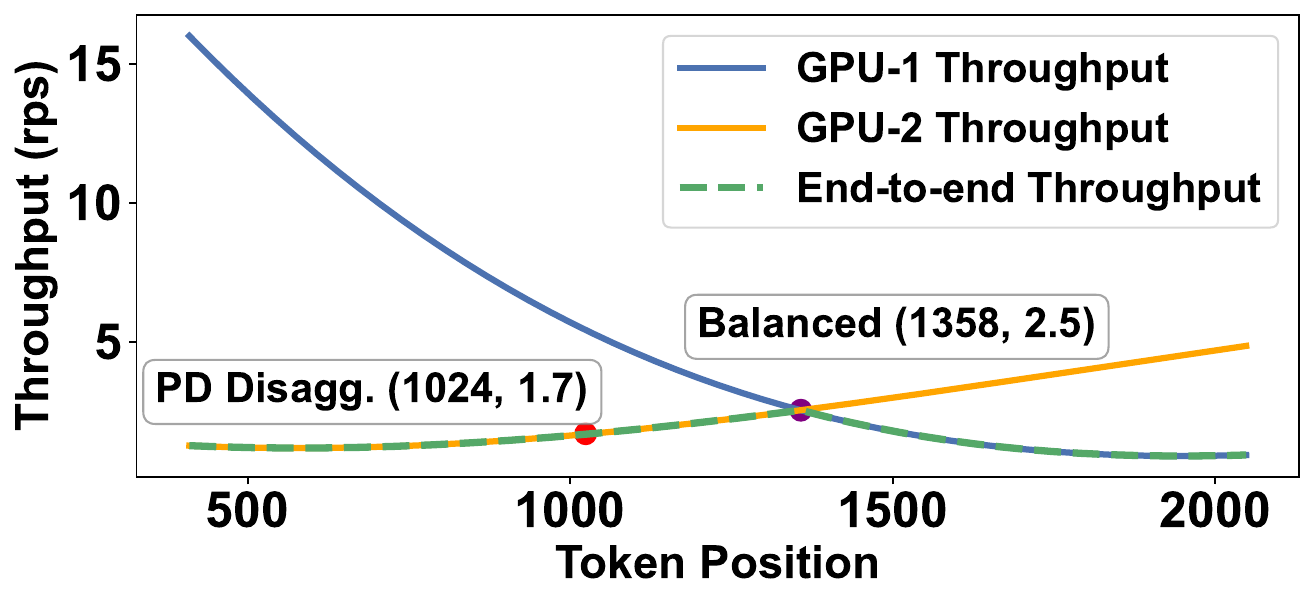}
    \caption{Throughput of Qwen2.5-32B on A100 under different split positions. Each request has 1024-token prompt and output. Position 1024 corresponds to PD disaggregation, while position 1358 represents the optimal split found by dynamic partitioning.}
    \label{fig:workload-hand-tune}
\end{figure}

To better guide the design of the partition policy, we first analyze single request partitioning in terms of performance. We conducted a controlled micro-benchmark using a synthetic workload, where each request consists of a fixed 1024-token prompt and a 1024-token output. The system runs on two A100 GPUs. We vary the partition position, which refers to the token index at which the request is split between the two GPUs, and observe its impact on throughput.

In~\autoref{fig:workload-hand-tune}, a partition position of 1024 corresponds to vanilla coarse-grained PD disaggregation: the entire prompt is handled by GPU-1, and the entire decode phase by GPU-2. As the plot shows, this static partitioning results in highly unbalanced workloads. GPU-1, responsible only for prefill, processes requests much faster than GPU-2, which is burdened with the full decode workload. Because the two GPUs operate as a pipeline, the overall system throughput is bottlenecked by the slower of the two—GPU-2 in this case.

We then gradually shift the partition point forward, incrementally assigning more decode tokens to GPU-1. As we do this, the throughput of GPU-1 decreases (due to the added decode load), while the throughput of GPU-2 increases (as its workload lightens). At a specific partition point (around PD ratio = 0.3), the two GPUs reach a balanced state where their processing rates align. At this point, the system achieves its maximum request throughput, since neither GPU is idle nor overloaded. This experiment shows that \textbf{Insight 1: maximizing system throughput depends on balancing request throughput or execution time across GPUs}.

While splitting a single request into two micro-requests with equal estimated execution time may seem balanced in isolation, it does not guarantee balanced load across GPU instances. In real-world settings with many concurrent requests, their micro-requests may be routed to the same GPU, causing contention and skewed resource usage due to overlapping execution and varying runtime behaviors. As a result, balanced per-request partitioning can still lead to system-wide imbalance. To address this, \sys extends its partitioning strategy to consider real-time loads of all GPU instances. Instead of treating each request in isolation, the global scheduler adjusts split decisions based on current system state, i.e., assigning fewer tokens to heavily loaded GPUs and more to underutilized ones.

\begin{algorithm}[!t]
\caption{\textsc{Global Request Scheduler}\,$(r)$ \small}
\label{alg:global}
\small
\KwIn{$r=(P,D,L,K)$ \hfill\textit{// prompt, predicted-decode, current server load, maximum step for binary search}}
\KwOut{partition ratio $\phi$}

\BlankLine
\If(\tcp*[f]{cold start}){$\mathsf{prefill\_clk}=0\;\land\;\mathsf{decode\_clk}=0$}{
  \Return \textsc{ColdStart}\,$(r)$
}\label{alg1 line2}

\medskip
\tcp{start from PD disaggregation}
$\phi\leftarrow \dfrac{P}{P+D}$, $lo\leftarrow 0$, $hi\leftarrow 1$\;\label{A1 line4}

\For{$k=1$ \KwTo $K$}{
  $(r_1,r_2)\leftarrow\textsc{Split}(r,\phi)$\;
  \tcp{total execution time on each server}
  $T_1,T_2\leftarrow\textsc{Predict}(r_1,r_2,L)$\;\label{A1predict}
  \If{$|T_1-T_2|\le\varepsilon$\label{A1 line11}}{
    \textbf{break}\tcp*[f]{execution time balanced}
  }
  \Else{
    \textsc{update}$(lo,hi,\phi)$
    \tcp*[f]{binary search update} 
  }
}

\textsc{Commit}$(r_1,r_2)$\;
\Return $\phi$
\end{algorithm}

\para{Global-level decision and request planner.}
Based on the above insight, the global scheduler must choose an appropriate partition ratio
\(\phi\in[0,1]\) per request for a stream of requests $\mathcal{R}=\{r_i\}$, to maximize the overall performance. As discussed in~\autoref{sec:overview}, the first \(\lceil\phi L\rceil\) tokens of the
micro-requests $\alpha$ are assigned to the \servera, the remainder to the \serverb. The sole objective is to make sure neither GPU idles while waiting for the other.

The \autoref{alg:global}  achieves this goal with a bounded binary
search. It starts from an initial ratio \(\phi = \frac{P}{P + D}\), which corresponds to pure PD disaggregation (\autoref{A1 line4}). In each iteration, the scheduler splits the incoming request \(r\) into two parts according to \(\phi\), and uses an analytical latency predictor to estimate the total execution time on each server, denoted by \(T_1\) and \(T_2\) (\autoref{A1predict}).
Here, \(T_1\) and \(T_2\) represent the predicted time for each server to complete all assigned micro-requests, including the new split segments, under the current load conditions. This prediction is derived from offline profiling and performed via efficient table lookups with a small LRU cache, so each probe costs only a few microseconds.
The binary search adjusts \(\phi\) to balance the workload by minimizing the difference between \(T_1\) and \(T_2\). The search stops once the absolute difference \(|T_1 - T_2|\) is within a small tolerance \(\varepsilon\) (\autoref{A1 line11}), indicating near-equal execution times on both servers. To limit overhead, the maximum number of binary search iterations is capped by the parameter \(K\) (typically set to 6 in our environment).

A special cold-start path seeds the prefill and decode clocks when the
first request arrives (\autoref{alg1 line2}), after which every new request reuses the
already-executed prefix of the timing arrays and simulates only the
delta it would add.  With constant-sized data per probe, the
per-request complexity is $O(1)$—no more than six simulator calls on our
cluster—yet the achieved throughput stays within one percent of an
offline oracle that has perfect future knowledge.

\para{Execution predictor within global scheduler.}
To gauge the benefit of a partition plan on two selected GPUs, the global scheduler uses a lightweight predictor that analytically estimates the plan’s timing. The predictor maintains an input queue of micro-requests and simulates a virtual batch under the same hardware constraints as the runtime: a per-pass batch composition drawn from each instance’s recent batching history, at most $N_{\max}$ concurrent requests, and the requirement that every active request advances by at least one token per pass. To preserve accuracy, the predictor continuously calibrates itself with real-time GPU metrics and recent execution progress, ensuring its model of system status remains up to date before predicting batch latency.

During each virtual pass, the predictor admits requests until the token budget or request limit is reached. It grants one prompt chunk or one decode token per request, then advances time. Advancing deducts granted tokens from remaining work, resets the token budget, and stores a compact batch snapshot including request IDs, outstanding tokens, and timestamps. This cycle repeats until all work is complete, producing a detailed timeline of future GPU activity.

First-Come-First-Serve is the default admission policy as it mirrors our production runtime, but the predictor is policy-agnostic. \sys can be adated to priority, shortest-job-first, or deadline-aware policies.  Because the predictor’s state snapshots are tiny and its computations are purely arithmetic, it adds negligible overhead while giving the Global Scheduler an accurate basis for tuning the split ratio $\phi$ and other placement decisions.

We use First-Come-First-Serve as the default admission policy to match vLLM setup.
The predictor adds negligible overhead, as it operates on lightweight state snapshots and uses only simple arithmetic, while assisting the global scheduler in tuning the partition ratio $\phi$.

\subsection{Local: SLO-Aware Batch Composition}
\label{sec:localsched}
\label{sec:ana_trade}

\begin{figure}[!tb]
    \centering
    \includegraphics[width=1\linewidth]{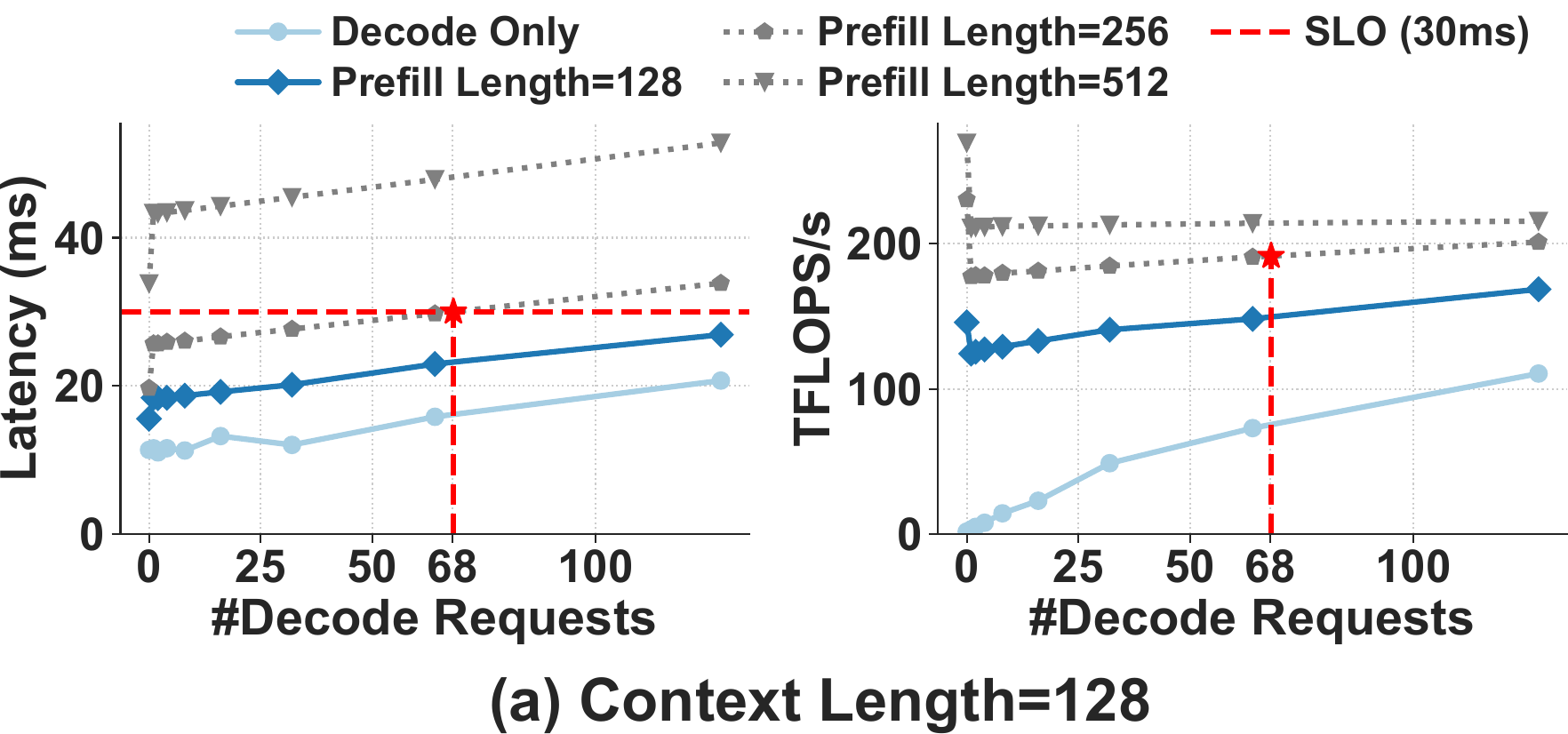}
    \includegraphics[width=1\linewidth]{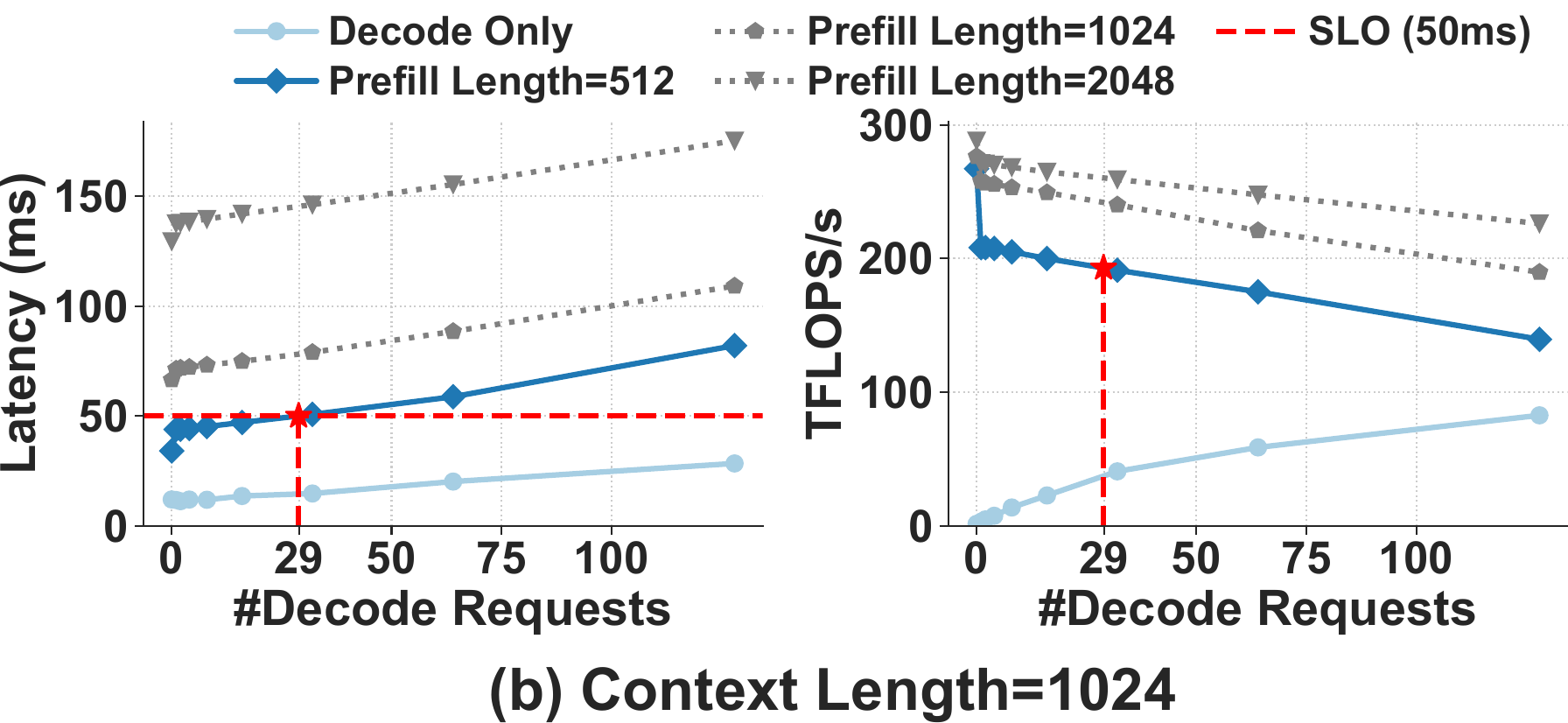}
    \caption{Latency and GPU compute utilization under different batching strategies for Llama-3.1-8B on an A100 GPU. Dashed red line represents latency SLO and the number of decode requests being processed. With a fixed decode workload, different sizes of prefill chunks can be mixed in a batch.}
    \label{fig:sec4.batch.latency}
\end{figure}
Complementing the global scheduler, the local scheduler manages batching on each GPU to ensure SLO compliance.

\para{Mix batch's performance analysis.}
\autoref{fig:sec4.batch.latency} tracks how the mixture of prefill and decode requests in a single batch affects both latency (left-hand figures) and GPU throughput in TFLOPs/s (right-hand figures) on an A100 running Llama-3.1-8B. Two context lengths are examined: a short context of 128 tokens (top row) and a long context of 1024 tokens (bottom row). In every latency figure the red dashed horizontal line marks the latency service-level objective, 30 ms for the short context and 50 ms for the long one. Where that line intersects a latency curve we define the Latency-Constrained Utilization (LCU) point. The x-coordinate of the LCU point reveals the largest number of concurrent decode requests that can be handled without breaching the SLO, while the corresponding throughput is obtained by reading straight up to the matching TFLOPs/s curve in the plot on the right. Because varying the prefill batch size shifts the position of the LCU point, the batch composition directly determines how efficiently real-time inference workloads can be served. From these results, we further derive the following key insights:

\begin{algorithm}[!t]
\caption{\textsc{Local Scheduler}\,$(P,\,D,\,S,\,B\,,T)$}
\label{alg:dcr}
\small
\KwIn{$P$, $D$, $S$, $B$, $T$ \hfill\textit{// prefill queue, decode queue, target SLO, previous batch, profile table}}
\KwOut{next batch $B$}
\BlankLine

$\textsc{Record(T, B.plen, B.ctx, B.dnum, B.time)}$\;

$B \leftarrow D $, $M \leftarrow \textsc{MaxPrefillAllowed}(T, S, B.ctx, B.dnum)$\;\label{alg2 line2}

\ForEach{$r \in P$}{\label{alg2 line3}
  \tcp{schedule no more than M tokens}
  $t = min(\textsc{r.Token}, M)$\; \label{line4}
  $B.add(r,t)$\; 
  \tcp{update budget}
  $M \leftarrow M - t$\;
    \If{$M \leq 0$}{
      \textbf{break}
    }
}

\Return $B$
\end{algorithm}

\para{Insight 2: decode–prefill trade-off between latency and GPU utilization.} Decode-only batches consistently meet the 50 ms SLO but are memory-bound, leaving much of the GPU’s compute capacity idle and yielding modest TFLOP/s throughput. Introducing a moderate prefill segment (e.g., 512 tokens) shifts work toward compute-bound operations, raising utilization and throughput; however, as the number of concurrent decode requests grows (for example, beyond 29 when decode length is 1024), the end-to-end latency went past the 50 ms target. Thus, higher utilization is achieved at the cost of latency increasing.

\para{Insight 3: batch composition (prefill length (plen), context length (ctx) and decode token number (dnum)) is key performance drivers—but the optimal configuration is dynamic.}
Larger prefill batches (e.g., 1024 tokens) can boost throughput in light‐load scenarios but quickly breach SLOs as batch sizes grow. Longer decode sequences further increase memory contention and delay phase handoffs. Throughput also depends on context length: short contexts (e.g., 128 tokens) exhibit a roofline pattern, rising with more decode requests until the LCU point. And long contexts (e.g., 1024 tokens) see throughput degrade under heavier decode loads. Here the LCU Point offers a concrete metric to evaluate throughput under latency constraints. These patterns suggest that request partitioning and scheduling are crucial for handling diverse prompt lengths and dynamic loads to achieve high GPU utilization without violating SLOs.

\para{Dynamic batch composition.}
Based on the above insights, we design our local scheduler. After the request is split, the local inference engines need to process them. A naive scheduling policy would simply prefill a fixed batch of tokens and let the decode thread emit one token per request; the batch size then acts as a crude latency knob. In practice this is ineffective, because inference latency depends not only on the batch size but on the full batch composition—prefill length, average context length, and the number of concurrent decodes.

\begin{figure}[!t]
    \centering
    \includegraphics[width=0.48\textwidth, keepaspectratio]{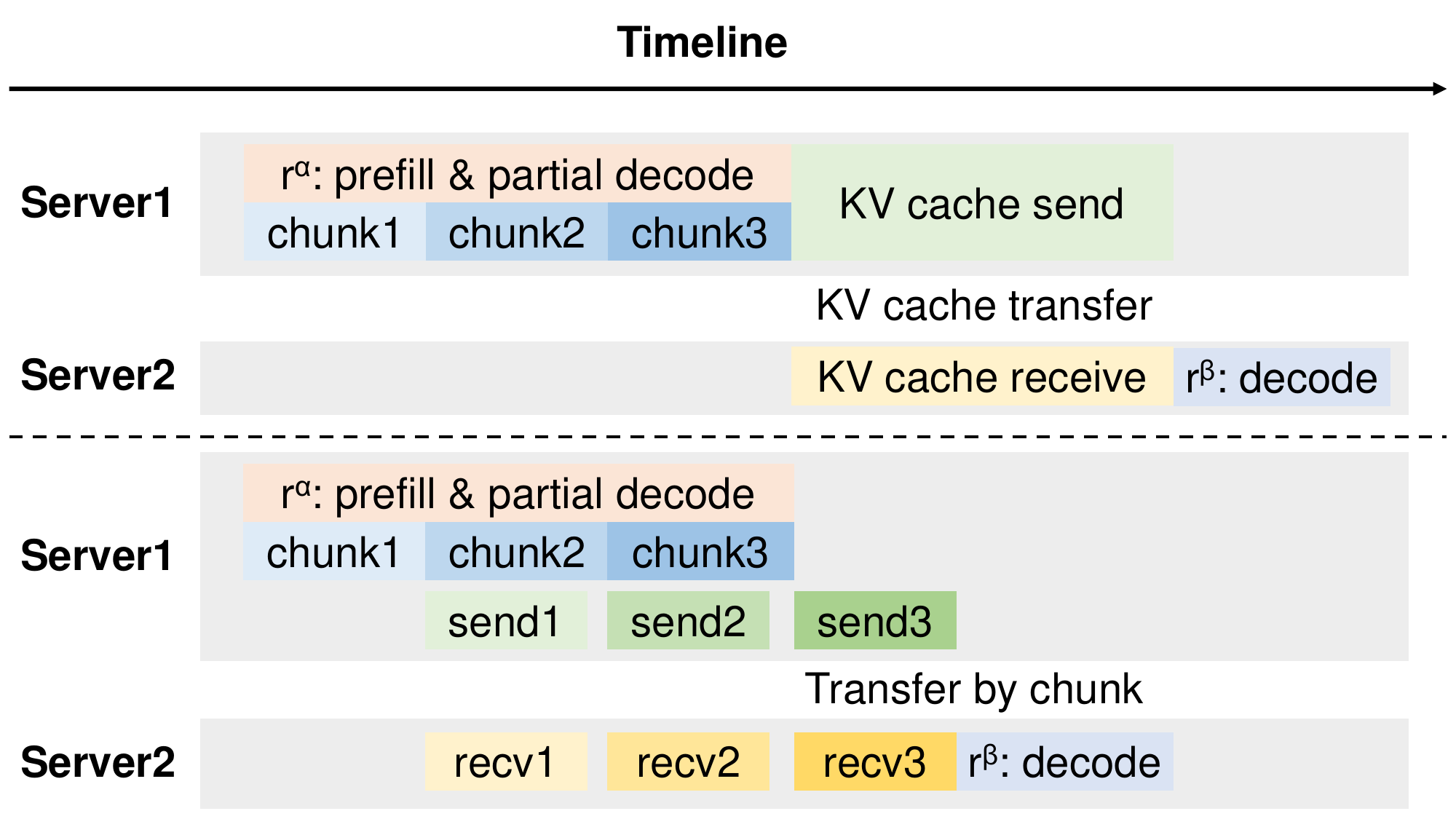}
    \caption{Transfer optimization implementation of \sys. The top part shows the original transfer transfer mechanism; the bottom part shows our own chunk-based transfer.}
    \label{fig:transfer_by_chunk_implementation}
\end{figure}

To make scheduling SLO-aware and hardware-efficient, we design a local scheduling algorithm that dynamically composes the batch before each hybrid kernel launch. As illustrated in Algorithm~\ref{alg:dcr}, the scheduler updates a profile table with latency statistics from the previously executed batch. Specifically, it records the tuple $(\textit{plen}, \textit{ctx}, \textit{dnum}, \textit{time})$, where \textit{time} is the measured latency. This step ensures that the profile table is progressively refined with execution feedback.

The scheduler then constructs the next batch in two stages. First, it includes all decode requests (\autoref{alg2 line2}), which are assumed to be latency-critical and must be processed immediately. Based on the decode portion of the batch, it computes the average context length and decode count. These values are used to consult the profile table and determine the maximum prefill token budget $M$ that would keep the batch latency within the target SLO.

Next, the scheduler traverses the prefill queue in arrival order and greedily adds as many requests as possible into the batch without exceeding the token budget $M$ (\autoref{alg2 line3}). Each prefill request $r$ may contribute up to $\min(r.\textsc{Token}, M)$ tokens (\autoref{line4}), and the budget $M$ is updated accordingly. The traversal stops once the budget is exhausted. This budget-aware selection maximizes utilization while preserving SLO.

This procedure, which records execution statistics, bounds prefill length by the SLO, and schedules each batch carefully, allows \sys to adaptively shape each batch based on runtime behavior and workload conditions.

\subsection{Runtime Support: Chunk-Based KV Transfer}
\label{sec:chunk-transfer}
\autoref{fig:transfer_by_chunk_implementation} illustrates how \sys transfers the KV cache of $r^\alpha$ across servers at chunk granularity. Server1 processes $r^\alpha$ in equal-sized chunks, regardless of token type. Once chunk $k$ completes, its KV block is immediately DMA-pushed to Server2, while Server1 continues with chunk $k+1$. This is both safe and efficient because the KV cache is append-only~\cite{sun2024llumnix}; completed chunks are immutable and can be transferred in parallel without coherence concerns. Transfers are fully offloaded to high-speed libraries like NCCL~\cite{nccl-lib} or Mooncake~\cite{qin2025mooncake}, with lightweight ZeroMQ~\cite{zmq} messages steering placement on the receiver side.
This chunk-level transfer enables \sys to ship KV blocks as soon as they are ready, overlapping communication with computation and supporting a unified interface for all micro-requests. While prior approaches operate at the layer or iteration level~\cite{qin2024mooncake,sun2024llumnix}, our method achieves finer granularity and can be composed with those techniques.

\section{Implementation and Discussion}

\noindent We implemented \sys on top of vLLM with 3K lines of Python code, as an end-to-end distributed LLM serving architecture featuring a two-level scheduler and backend integration interface. The design is non-intrusive and can be extended to various other inference engines. We will open source our code in the near future.

\para{Request scheduler.}
The global scheduler in \sys is integrated into the front-end proxy of \texttt{vLLM}. On arrival, each request is passed through a length predictor, then routed to the scheduler, which chooses a partition ratio $\phi$ and dispatches the resulting micro-requests to a pair of GPU engines. A lightweight latency predictor (\autoref{sec:profiler}) embedded in the controller estimates per-engine throughput in real time. To reduce scheduling overhead, we implemented the global scheduler logic in C++ while maintaining full compatibility with the Python-based codebase.

\para{Prediction length discussion.}
\sys uses predicted output lengths to guide request partitioning but remains robust under moderate inaccuracies. In our experiments, over 95\% of predictions fall within $\pm 100$ tokens of the actual value, and large deviations are rare. This accuracy is consistent with prior work~\cite{jin2023s}. The global scheduler bases decisions on relative execution time across GPU stages rather than exact token counts, so small prediction errors have minimal impact. To improve robustness, we add a configurable margin (20 tokens in our setup) to avoid underestimation, which could lead to SLO violations. Slight overestimation results in only minor efficiency loss. 
Our sensitivity analysis (\autoref{tab:predict_sens}) confirms that \sys sustains performance targets across realistic prediction errors.

%% file: sec/eval.tex
\section{Evaluation}

\subsection{Experimental Setup}

\para{Models.} We evaluate \sys using three open-source LLM models from the popular Qwen-2.5 series~\cite{qwen2025qwen25technicalreport}, including Qwen-2.5-14B~\cite{qwen-14b}, Qwen-2.5-32B~\cite{qwen-32b}, and Qwen-2.5-72B~\cite{qwen-72b}. 

\para{Testbed.} Our evaluation platform consists of two servers, each with four NVIDIA A100 80GB GPUs, 128 CPUs, 1TB RAM, and 4$\times$200 Gbps ConnectX-6 RoCE NICs.  NVLink provides 600 GB/s bandwidth between any two GPUs within a server. We use PyTorch 2.5.1 and CUDA 12.4 throughout our experiments. Unless otherwise specified, all evaluations are conducted on two servers.

\para{Workloads.} We generate request arrival patterns using a Poisson distribution, as in prior work~\cite{kwon2023efficient,wu2024loongserve,zhong2024distserve}. Request input and output lengths are sampled from multiple real-world datasets, including Azure Code~\cite{azuredata}, BurstGPT~\cite{wang2024burstgpt}, arXiv Summarization~\cite{cohan-etal-2018-discourse}, and Mini Reasoning~\cite{mini-reasoning}. These diverse settings reflect a variety of prompt/output length distributions and task complexity.

\begin{figure*}[!t]
    \centering
    \includegraphics[width=0.9\textwidth]{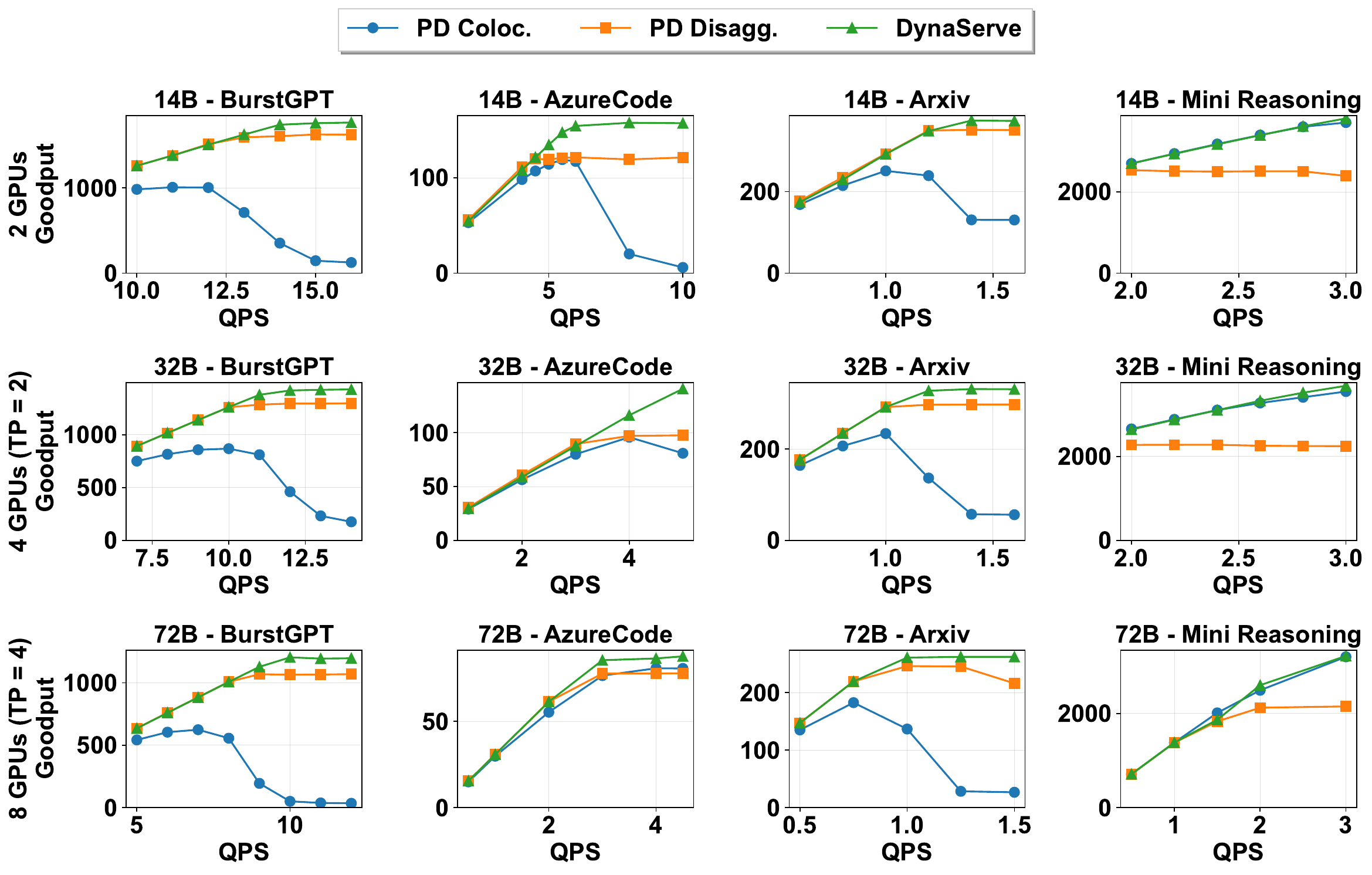}
    \label{cap:d2p}
    \caption{Goodput of \sys, PD Colocation, and PD Disaggregation under a 100\,ms TBT SLO. Rows represent model sizes (14B, 32B, 72B); columns represent workloads (BurstGPT, AzureCode, arXiv Summarization, and Mini Reasoning).}
    \label{fig:goodput_overall}
\end{figure*}

\para{Baselines.} We compare \sys against the state-of-the-art LLM serving system vLLM (v0.7.2) in two configurations. \textbf{PD Colocation (PD Coloc.)}: We use vLLM’s default chunked prefill strategy, with chunk sizes tuned between 256–2048 for each workload. Since Sarathi-Serve's~\cite{agrawal2024taming} Chunked Prefill has already been adopted by vLLM, we do not evaluate Sarathi-Serve separately; \textbf{PD Disaggregation (PD Disagg.)}: We extend vLLM’s original disaggregation mode (v0) to a more advanced version (v1) under modern scheduling logic. DistServe~\cite{zhong2024distserve} shares a design very similar to vLLM’s PD mode; hence, we omit an independent comparison.
We focus on these baselines because they represent common design patterns, and reusing vLLM ensures fair comparisons. 

We deploy models using data (DP) and tensor parallelism (TP). For PD Coloc., we use 2 GPUs (DP=2) for 14B, 4 GPUs (TP=2, DP=2) for 32B, and 8 GPUs (TP=4, DP=2) for 72B.
For PD Disagg., we use TP within separate prefill and decode instances: we use 1P1D for 14B, 2P2D for 32B, and 4P4D for 72B. \sys uses the same number of GPUs as PD Disagg., allocating TP groups to process $r^\alpha$ and $r^\beta$ separately: 1 GPU each for $r^\alpha$ and $r^\beta$ in 14B (2 GPUs total), 2 GPUs each in 32B (4 GPUs total), and 4 GPUs each in 72B (8 GPUs total).

\para{Metrics.} We evaluate \sys using two primary metrics: \textbf{goodput}~\cite{zhong2024distserve}, defined as the number of tokens generated per second while meeting service-level objectives (SLOs), and \textbf{serving capacity}, defined as the maximum sustainable queries-per-second (QPS) under SLO constraints, following the definition used in Sarathi Serve~\cite{agrawal2024taming}. We enforce a strict 100~ms time-between-tokens (TBT) SLO to ensure responsive interactive performance, following prior work~\cite{zhong2024distserve}. This 100~ms target, commonly adopted in practice, is applied uniformly across all model sizes. Since larger models are configured with higher TP degrees, they are expected to meet the same latency budget despite their increased computation.

\subsection{Overall Results}

\autoref{fig:goodput_overall} compares the goodput of \sys, PD Coloc., and PD Disagg. across various workloads and model scales. \sys consistently outperforms both baselines across all evaluated settings. Specifically, it achieves a maximum goodput improvement of up to 91\% over PD Coloc., and up to 61\% over PD Disagg.. These gains reflect \sys's effective load balancing, reduce PD interference, and maintain stable performance under increasing QPS within the SLO.

In  BurstGPT, Azure Code, and arXiv Summarization workloads, \sys demonstrates a steady increase in goodput with rising QPS, until reaching saturation where it maintains a plateau. In contrast, PD Coloc. shows degraded performance beyond a certain QPS due to PD interference. This is mitigated in \sys by only introducing mixed PD batches when imbalance arises, thereby reducing the likelihood and severity of interference. Furthermore, the dynamic batch composition strategy helps constrain batch-level delays and maintain TBT within the SLO threshold. PD Disagg., while immune to direct PD interference, lacks scheduling flexibility due to its rigid prefill/decode separation.
As a result, it often suffers from server underutilization in skewed workload .

In the Mini Reasoning workload, which exhibits more pronounced imbalance due to longer generation lengths, \sys benefits substantially from its adaptive scheduling strategy. The ability to dynamically partition requests between servers improves utilization and results in higher goodput. Interestingly, in this particular workload, PD Coloc. also achieves relatively strong performance. This is because the prefill stage accounts for a smaller fraction within the workload, making decoding less sensitive to prefill-induced delays and PD Coloc. becoming a very strong baseline.

As model size grows from 14B to 72B, PD Coloc. suffers increasing PD interference, causing sharp goodput drops past peak QPS. In contrast, \sys maintains stable throughput and consistently outperforms PD Disagg., showing strong resilience to scaling and load.

\subsection{Serving Capacity}

\begin{figure}[!tb]
    \centering
    \includegraphics[width=0.47\textwidth]{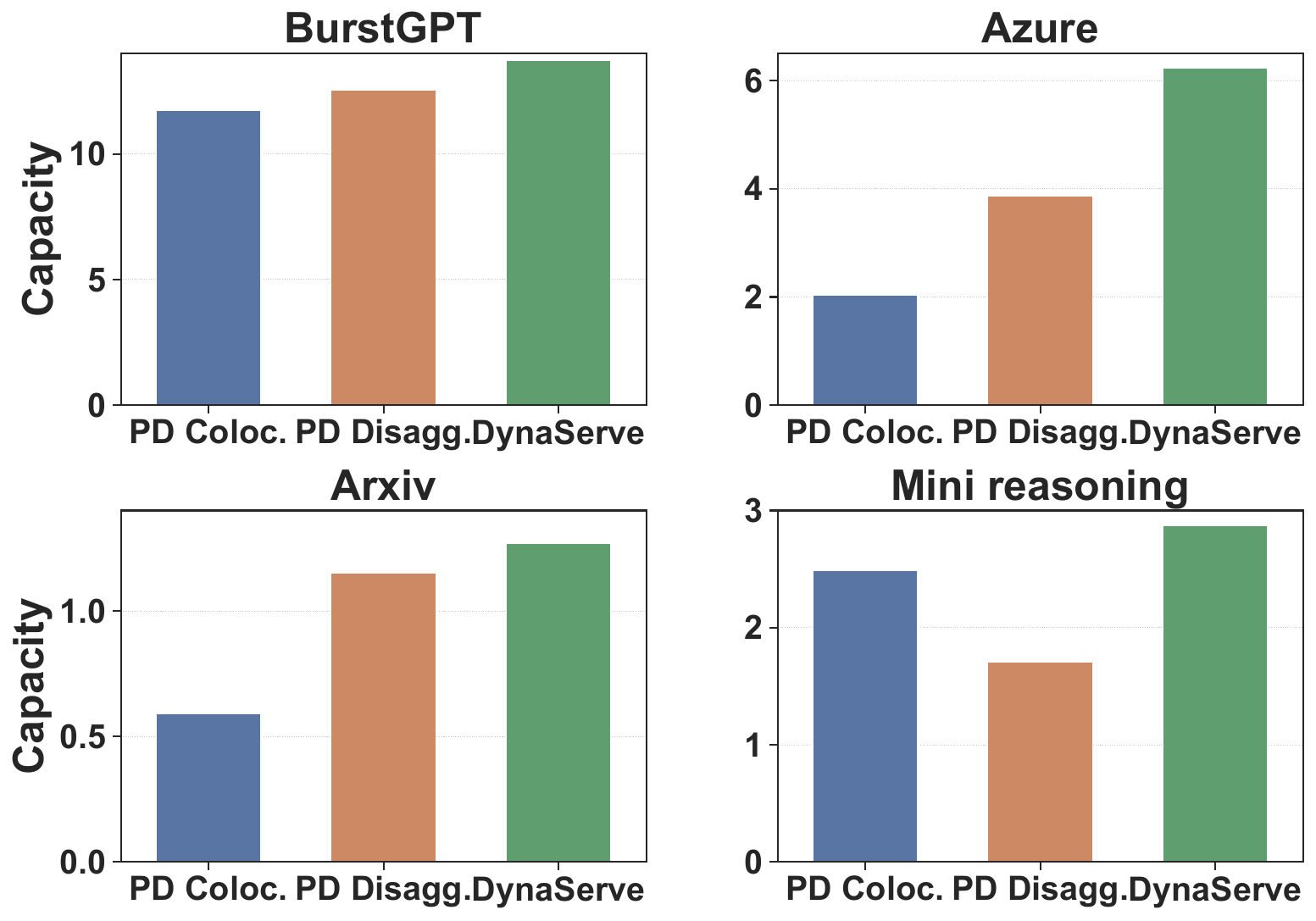}
    \label{cap:d2p}
    \caption{Serving capacity of \sys, PD Colocation, and PD Disaggregation across four workloads using the Qwen-14B.}
    \label{fig:capcity}
\end{figure}

In addition to goodput, we evaluate the maximum queries per second (QPS) each system can support while keeping the p99 token-by-token latency (TBT) under the 100\,ms SLO. This evaluation is conducted across four workloads using the Qwen-2.5-14B model. We refer to this metric as Serving Capacity, which is more SLO-strict-allowing only 1\% of requests to violate the TBT SLO. As shown in~\autoref{fig:capcity}, \sys consistently delivers the highest serving capacity. On average, \sys supports 2.37$\times$ the QPS of PD Coloc. and 1.37$\times$ that of PD Disagg., highlighting its strong adaptability and efficiency under diverse traffic patterns.

\sys follows a similar trend to goodput, thanks to its ability to dynamically balance compute and control interference. Overall, its elastic scheduling and adaptability across workload shapes enable higher capacity, being crucial for efficient large-scale LLM serving. For decode-light workloads like AzureCode, Disaggregation performs better than Colocation by isolating prefill traffic, but suffers decode underutilization due to coarse-grained partitioning. \sys overcomes both issues, achieving up to 3$\times$ higher QPS than Colocation. On decode-heavy workloads like Mini Reasoning, Colocation benefits from batching efficiency, yet \sys further improves capacity by dynamically controlling batch granularity and load distribution. Even for long-input workloads like arXiv Summarization, where Disaggregation reduces prefill contention, \sys still leads slightly by coordinating both phases more flexibly. On BurstGPT, where prefill and decode are balanced, \sys gains stem from mitigating interference without underutilizing servers.

\subsection{Hybrid Workload}

\begin{table}[!t]
\centering
\caption{Serving capacity under a hybrid workload (50\% BurstGPT + 50\% Azure Code) using the Qwen-14B model.}
\label{fig:hybird_work}
\resizebox{0.45\textwidth}{!}{
\begin{tabular}{@{}l@{\hskip 10pt}ccc@{}}
\toprule
\textbf{System} & \textbf{PD Coloc.} & \textbf{PD Disagg.} & \textbf{\sys} \\
\midrule
\textbf{Serving Capacity (rps)} & 4.6 & 5.9 & 7.4 \\
\textbf{Goodput (token/s)}   & 316.32 & 399.31 & 473.84 \\
\bottomrule
\end{tabular}
}
\end{table}

To reflect real-world usage where request patterns span diverse tasks, we construct a hybrid workload by uniformly mixing BurstGPT and Azure Code requests. This combination introduces contrasting prompt and response characteristics, making static partitioning inherently unbalanced.

As shown in~\autoref{fig:hybird_work}, \sys achieves 60\% higher serving capacity than PD Coloc. and 25\% higher than PD Disagg.. In terms of goodput, \sys outperforms PD Coloc. by 49\% and PD Disagg. by 20\%. PD Coloc. suffers from interference, as large Azure prompts consume compute and degrade latency across both workloads. PD Disagg. avoids interference but cannot fully utilize resources due to fixed partitioning, which fails to serve both workloads efficiently. In contrast, \sys dynamically adjusts the partition ratio per request, balancing compute and memory usage across GPUs. This adaptivity leads to better utilization, stable latency, and higher overall capacity for \sys.

\subsection{Real-time Workload Replay}

\begin{figure}[!tb]
    \centering
    \includegraphics[width=0.43\textwidth]{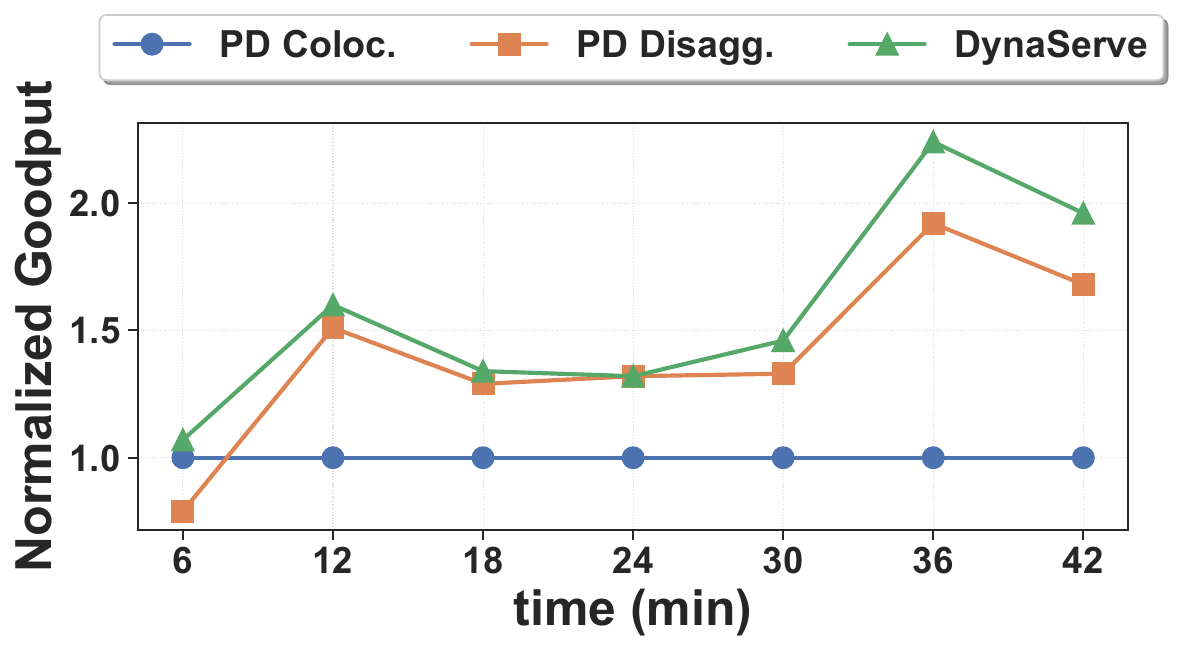}
    \label{cap:d2p}
    \caption{Goodput trends under a real-time workload from the BurstGPT dataset, using the Qwen-14B model.} 
    \label{fig:realtime}
\end{figure}

To evaluate performance under realistic temporal dynamics, we extract a continuous request stream from the BurstGPT dataset, preserving original arrival pattern to simulate a realtime workload. The extracted trace starts at 311th hour and spans 42 minutes, and we measure average goodput every 6 minutes. 
\autoref{fig:realtime} shows goodput over time for all systems. We observe fluctuations in the measured goodput, which are primarily attributed to variations in input lengths. During intervals with longer prompts, the system typically exhibits reduced goodput, resulting from increased prefill overhead. 
During the time interval between 12 and 42 minutes, PD Disaggregation outperforms PD Colocation, thanks to its isolation of prefill and decode stages which mitigates interference. \sys, however, further improves goodput by employing \techls to limit interference and \techgs to dynamically rebalance load between GPU instances, achieving consistently higher throughput.

In contrast, during the first 6 minutes, requests are more decode-heavy with relatively short prefill phases. Here, Colocation suffers less from interference and temporarily surpasses Disaggregation. Even in these cases, \sys leverages its flexible \techgs mechanism to maintain top-tier goodput across the board.

\subsection{Breakdown Analysis}

\begin{figure}[!tb]
\centering
\includegraphics[width=0.45\textwidth]{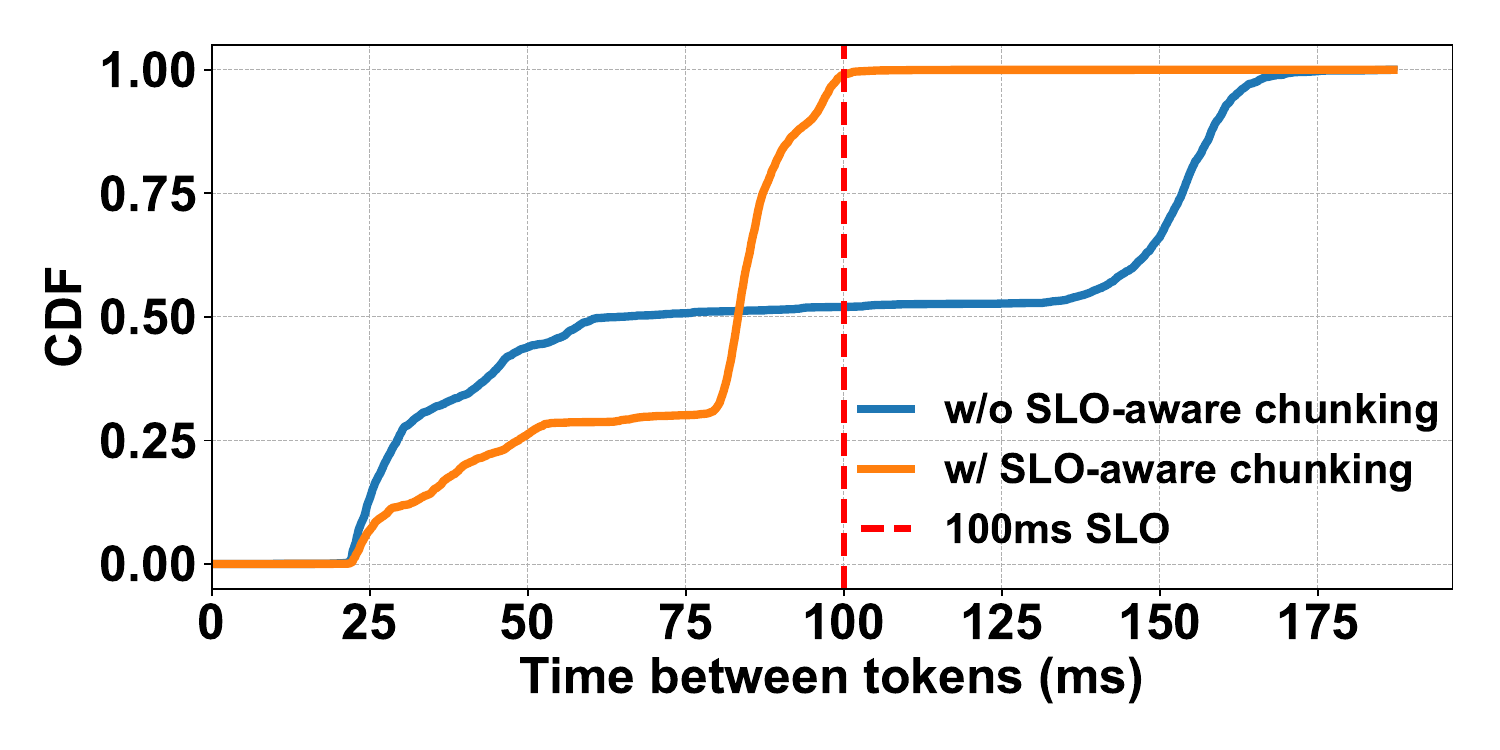}
\caption{CDF of time-between-tokens (TBT) with and without \techls, using Qwen-14B model and AzureCode workload. QPS is set to serving capacity of \sys.}
\label{fig:itl_cdf}
\end{figure}

Then we dive into \sys to investigate its internal details.

\para{Benefit of \techls.}
We first evaluate the impact of \techls by measuring the CDF of TBT under high-workload pressure conditions. As shown in~\autoref{fig:itl_cdf}, without \techls, PD interference leads to significant tail latency, with TBT exceeding 175\,ms in the worst case. Only 52\% of tokens are generated within the 100\,ms SLO threshold, meaning nearly half the tokens violate the service requirement. By enabling \techls, \sys effectively constrains PD interference and tail delay, raising the SLO-attainment to 99\%. This demonstrates the importance of fine-grained batch adjustment in ensuring latency predictability under mixed prefill-decode batches.

\begin{table}[!tbp]
\centering
\caption{Per-request scheduling overhead under varying QPS when serving Qwen-14B with BurstGPT traces.}
\label{tab:scheduling_overhead}
\resizebox{0.48\textwidth}{!}{
\begin{tabular}{@{}lcccccc@{}}
\toprule
\textbf{QPS} & \textbf{6} & \textbf{8} & \textbf{10} & \textbf{12} & \textbf{14} & \textbf{16} \\
\midrule
\textbf{Overhead (ms)} & 17.47 & 15.48 & 14.40 & 13.70 & 14.85 & 14.54 \\
\bottomrule
\end{tabular}
}
\end{table}

\para{Scheduling overhead analysis.} Secondly, we measure the runtime overhead of the global scheduler in \sys. We evaluate this using the Qwen-14B model in a 2 GPUs setup (1 for $r^\alpha$ and 1 for $r^\beta$), replaying BurstGPT traces under varying QPS levels. As shown in \autoref{tab:scheduling_overhead}, the per-request scheduling overhead remains consistently below 20~ms across all tested workloads. Since each request is scheduled only once at arrival, this one-time cost is amortized over the entire request lifetime. Given that typical end-to-end latency is around 5000~ms at 10 QPS, the overhead is negligible.

\begin{table}[!tbp]
\centering
\caption{Sensitivity of goodput to prediction errors. The scheduler assumes an output length of 1467 tokens, while actual output lengths are sampled from a normal distribution with varying standard deviation. Prompt length is 219 tokens.}
\label{tab:predict_sens}
\resizebox{0.47\textwidth}{!}{
\begin{tabular}{@{}l@{\hskip 10pt}cccc@{}}
\toprule
\textbf{Standard deviation
 ($\sigma$)} & \textbf{0} & \textbf{10} & \textbf{50} & \textbf{100} \\
\midrule
\textbf{Goodput (token/s)} & 3606.93 & 3591.00 & 3564.90 & 3501.85 \\
\bottomrule
\end{tabular}
}
\end{table}

\para{Sensitivity analysis.} Then, we evaluate \sys's robustness to output length prediction errors. We conduct a sensitivity analysis where the global scheduler assumes a fixed output length of 1467 tokens. Actual output lengths are drawn from a normal distribution with the same mean (1467) and varying standard deviations ($\sigma = 0, 10, 50, 100$). The prompt length is fixed at 219. As shown in~\autoref{tab:predict_sens}, \sys maintains high goodput across all levels of variance, with only a 2.9\% drop at $\sigma = 100$, where 95\% of output lengths fall between 1267 and 1667 tokens. These results demonstrate \sys's tolerance to moderate prediction error. 

\para{Chunk-based KV transfer.} Finally, we test chunk-based KV transfer using the Mini-Reasoning task as an example. Compared to a version of \sys without this technique, the chunked approach reduces non-overlapped transfer by 94\%, demonstrating its effectiveness in overlapping communication.

%% file: sec/relate.tex
\section{Other Related Work}

\noindent Optimizing Large Language Model (LLM) inference involves various strategies aimed at enhancing throughput and reducing latency. Recent works focused on multiple granularities.

\para{Request-level scheduling.} Latency–throughput tradeoffs are central to LLM serving. Orca~\cite{yu2022orca} improves throughput via continuous batching, while Apparate~\cite{dai2024apparate} reduces latency using dynamic early-exit points. Nanoflow~\cite{zhu2024nanoflow} overlaps communication and computation within nano-batches for better intra-device parallelism. NIYAMA~\cite{goel2025niyamabreakingsilos} adapts chunk sizes to support multi-SLO and priority workloads. These techniques offer advanced latency control for diverse scenarios. \sys is compatible with and can benefit from such methods, incorporating their insights into its own scheduling.

\para{Phase-level scheduling.} Apart from DistServe~\cite{zhong2024distserve}, Splitwise~\cite{patel2024splitwise} also disaggregates prefill and decode across clusters to reduce interference. Several recent works aim to improve resource utilization under PD disaggregation. Shubha~\cite{shubha2024usher} extends phase-level disaggregation with heterogeneous GPU support, speculative decoding, and dynamic scheduling. MLC-LLM~\cite{mlc-llm} statically offloads prefill to decode instances using a fixed ratio, without supporting decode migration. Semi-PD~\cite{hong2025semipdefficientllmserving} uses SM-level control for intra-GPU sharing between prefill and decode. Adrenaline~\cite{liang2025injectingadrenalinellmserving} offloads decode attention kernels to prefill instances to improve memory efficiency. In contrast, \sys introduces flexible, request-level scheduling that supports both merging partial decode with prefill and offloading partial prefill to decode, while remaining compatible with intra-GPU sharing techniques.

%% file: sec/conclusion.tex
\section{Conclusion}
\noindent We introduced \sys, a scalable and adaptive LLM serving system for dynamic, imbalanced workloads. \sys decomposes requests into micro-requests and schedules them across GPUs using a two-level scheduler. The global scheduler selects split points based on predicted latency and GPU loads, while local schedulers adjust batch composition to meet strict SLOs. This unified execution model combines the strengths of colocation and disaggregation, achieving high utilization and low tail latency.